# Oxygen vacancy-induced anomalous Hall effect in a nominally non-magnetic oxide

Athby H. Al-Tawhid[1], Jesse Kanter[2], Mehdi Hatefipour[2], Douglas L. Irving[1,3], Divine P. Kumah[3], Javad Shabani[2], and Kaveh Ahadi[1,3,*]

[1]Department of Materials Science and Engineering, North Carolina State University, Raleigh, NC 27695, USA
[2]Center for Quantum Phenomena, Department of Physics, New York University, New York 10003, USA
[3]Department of Physics, North Carolina State University, Raleigh, North Carolina 27695, USA

[*] Corresponding author.  Email: kahadi@ncsu.edu

**ABSTRACT**

The anomalous Hall effect, a hallmark of broken time-reversal symmetry and spin-orbit coupling, is frequently observed in magnetically polarized systems. Its realization in nominally non-magnetic systems, however, remains elusive. Here, we report on the observation of the anomalous Hall effect in nominally non-magnetic $KTaO_3$. Anomalous Hall effect emerges in reduced $KTaO_3$ and shows an extrinsic to intrinsic crossover. A paramagnetic behavior is observed in reduced samples and confirmed using first principles calculations. The observed anomalous Hall effect follows the oxygen vacancy-induced magnetization response, suggesting that the localized magnetic moments of the oxygen vacancies scatter conduction electrons asymmetrically and give rise to the anomalous Hall effect. The anomalous Hall conductivity is found to be insensitive to the scattering rate in the low temperature limit ($T$<5 K), implying that the Berry curvature of the electrons on the Fermi surface controls the anomalous Hall effect. Our observations provide a detailed picture of many-body interactions, which trigger the anomalous Hall effect in a non-magnetic system.

The description of the anomalous Hall effect (AHE) of magnetically polarized systems[1] in terms of the Berry curvature of the quasiparticles on the Fermi surface is one of the major accomplishments of modern condensed matter theory[2–7]. The mobile charge carriers gain a transverse momentum in systems with a magnetic polarization and spin–orbit coupling. The built-up Hall voltage, i.e. anomalous Hall component, can be observed as a distinct, additional contribution in the Hall measurements, superposed on the ordinary Hall effect[8–13]. The AHE originates from spin-orbit coupling and is proportional to the magnetization, $R_{xy}^{AHE}(B) = \lambda M_z(B)$, where $\lambda$ relates magnetization to the anomalous Hall component[14]. Theoretical[15–17] and experimental[18–20] observations of defect-induced magnetization were previously reported in non-magnetic oxides. Experimental realization of anomalous Hall effect, however, remains elusive in these systems.

The degree of disorder within a system determines the nature of the AHE response. The intrinsic AHE occurs in the moderately dirty regime in which anomalous Hall conductivity becomes non-dissipative, i.e. independent of the scattering rate[21]. In this regime, the Berry phase of the quasiparticles on the Fermi surface acts as an effective magnetic field generating a transverse momentum. In the dirty regime, however, the extrinsic anomalous Hall conductivity, $\sigma_{xy}^{AHE}$, is directly proportional to the longitudinal conductivity, $\sigma_{xy}^{AHE} \propto \sigma_{xx}^{1.6}$, and asymmetric scattering of the conduction electrons gives rise to AHE. The extrinsic to intrinsic *crossover* occurs when the scattering rate becomes comparable to Fermi energy[22,23]. In the clean regime, an extrinsic anomalous Hall response emerges due to skew scattering of charge carriers and shows a linear relationship with longitudinal conductivity, $\sigma_{xy}^{AHE} \propto \sigma_{xx}$.

$KTaO_3$ is an incipient ferroelectric[24] in which a combination of a small effective mass of electrons and large dielectric constant, i.e. giant effective Bohr radius[25], induces a metallic state even in the dilute doping regimes. The $KTaO_3$ conduction states are derived from Ta 5$d$ states and have higher mobility and spin-orbit coupling compared to Ti 3$d$ states in $SrTiO_3$[26,27]. Spin-orbit coupling lifts the degeneracy of Ta 5$d$ states and splits them into $J=3/2$ and $J=1/2$ with a 0.4 eV energy gap, where $J$ is total angular momentum[28]. Real space and momentum space-induced topologically nontrivial Hall signals were recently reported in $KTaO_3$[29].

Here, we report on the observation of an anomalous Hall signal in reduced $KTaO_3$. An extrinsic AHE emerges at low temperature ($T<60$ K). Our observations suggest that oxygen vacancies play the role of paramagnetic point defects, scattering the conduction electrons and giving rise to AHE. The anomalous Hall conductivity becomes independent of the scattering rate below 5 K, suggesting an extrinsic to intrinsic AHE *crossover*.

Oxygen vacancies were introduced to KTaO$_3$ (100) single crystal substrates in ultra-high vacuum using a molecular beam epitaxy system. A 10 nm TiO$_x$ layer was deposited at 800 °C and 1×10$^{-9}$ Torr on the surface of KTaO$_3$. Here, the TiO$_x$ layer acts as an oxygen getter and capping layer during annealing. The substrate became dark due to introduction of oxygen vacancies after annealing. The temperature dependent magneto-transport measurements were performed using a Quantum Design physical property measurement system (PPMS) with a lock-in amplifier (SR830, Stanford Research Systems) in AC mode with an excitation current of 10 µA and a frequency of 13.33 Hz. Magneto-transport measurements were performed using a Van der Pauw configuration. Gold contacts were deposited using a sputter system at the corners of the samples through a shadow mask. The Hall experiments were carried out at various temperatures (Supplementary materials, S1). The temperature-dependent Hall carrier densities extracted, $n = -1/(eR_H)$, where $R_H$ is the Hall coefficient and $e$ is the elementary charge. The Hall coefficient, $R_H = dR_{xy}/dB$, was extracted from a linear fit to the transverse resistance at a magnetic field range where AHE is saturated. Magnetization measurements were performed using a Quantum Design Superconducting Quantum Interference Device (SQUID) magnetometer. Magnetization samples were handled specifically to avoid any magnetic contamination. Magneto-transport measurements below 1 K were carried out in a Triton, Oxford Instruments, bottom loading dilution refrigerator.

Oxygen vacancies introduce itinerant electrons to Ta 5*d* derived states. Figure 1(a) shows resistance with temperature (300-0.03 K). The sheet carrier density is 2.5×10$^{15}$ cm$^{-2}$ at 60 K and reveals carrier freeze out below 5 K, reaching 4×10$^{14}$ cm$^{-2}$ at 30 mK, Fig. 1(b). The carrier freezeout could be due to the presence of defect states, including oxygen vacancy defects, acting as traps[30]. The residual resistivity ratio ($\rho_{300\,K}/\rho_{0.03\,K}$) is 342, and the carrier mobility increases from ~20 cm$^2$/Vs at room temperature to ~9,000 cm$^2$/Vs at 30 mK due to screening of the longitudinal optical phonons. The charge carriers are distributed across the substrate thickness, creating a 3D electron system. The uniformity of induced electron system, however, is not clearly understood. All the conduction electrons at the anomalous Hall experiment temperatures are from *J*=3/2, Ta 5*d* states due to the large spin-orbit coupling gap (0.4 eV[26,28]). KTaO$_3$ shows a relatively large and positive longitudinal magnetoresistance at 30 mK as shown in Fig. 1(c). The large magnitude of magnetoresistance is consistent with the expected parabolic relation between longitudinal magnetoresistance and relaxation time of charge carriers[31]. This contrasts with the Kondo picture of dilute magnetism in which a negative longitudinal magnetoresistance is expected[32,33]. Longitudinal magneto-conductance shows signatures of weak anti-localization (Supplementary materials, S2). Only the 2D electron system at the $KTaO_3$ interface demonstrates large coherence length and signatures of weak anti-localization correction[34–36].

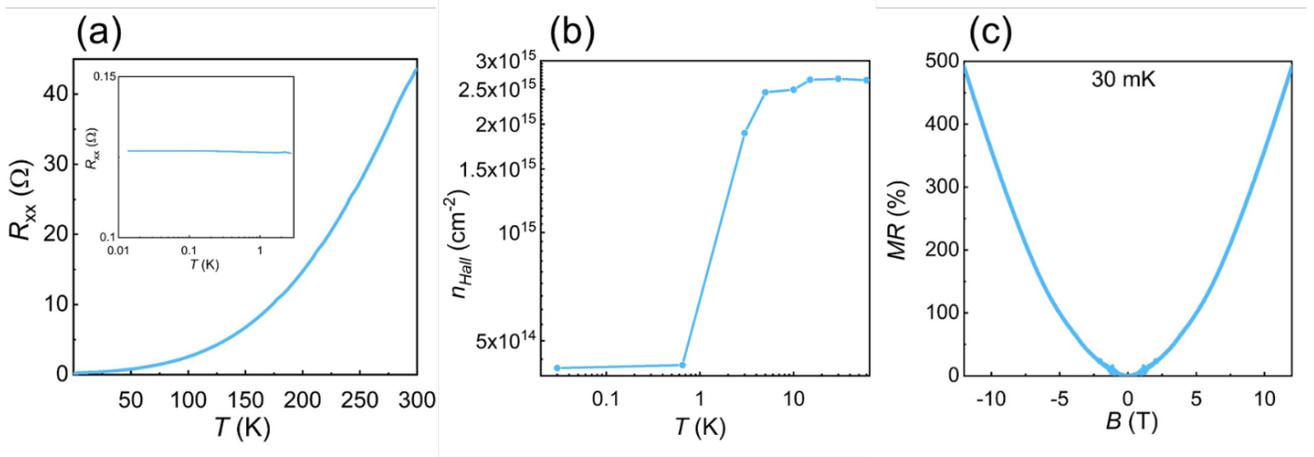

**Figure 1. Temperature- and magnetic field-dependent electronic transport.** (a) Resistance and (b) Hall carrier density in reduced $KTaO_3$ with temperature. The Hall carrier density shows a strong carrier freeze-out below 5 K. (c) Longitudinal magnetoresistance at 30 mK.

Figure 2 shows the AHE at different temperatures. The raw Hall results were anti-symmetrized to eliminate the longitudinal magnetoresistance contribution, $R_{xy} = [R_{xy}^{raw}(+B) - R_{xy}^{raw}(-B)]/2$. The ordinary Hall effect was subtracted from the anti-symmetrized Hall to resolve the AHE contribution, $R_{AHE} = R_{xy} - R_H B$. The AHE shows a monotonic increase and eventual saturation with the applied magnetic field. The details of AHE data extraction is described elsewhere[37–39]. A two-carrier fit to Hall measurement cannot explain the AHE response (Supplementary materials, S3).

The AHE saturation resistance ($R_{AHE}^{Sat.}$) does not change systematically with the temperature at higher temperatures (5-60 K) and shows an abrupt enhancement below 5 K, peaking at 30 mK ($20 \cong R_{AHE}^{Sat.}(0.03\ K)/R_{AHE}^{Sat.}(3\ K)$). The non-vanishing AHE suggests that the mechanism(s) leading to the AHE remain down to low temperatures. The abrupt enhancement of AHE coincides with a carrier freeze out (Fig. 1(b)) and suggests a potential change in the AHE mechanism.

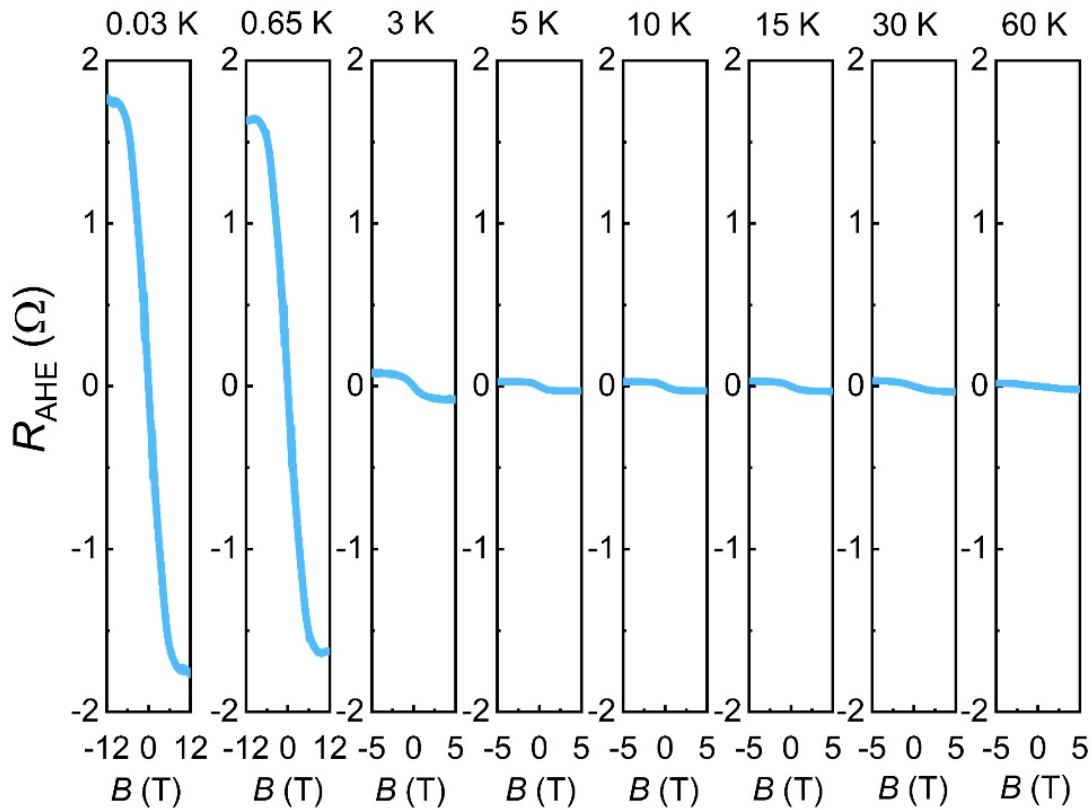

**Figure 2. Anomalous Hall effect in KTaO₃.** The extracted AHE after subtraction of the ordinary hall component in reduced KTaO$_3$ at different temperatures. The AHE resistance and saturation magnetic field increase below 5 K.

KTaO$_3$ is nominally non-magnetic, but AHE could arise when the localized paramagnetic defects, polarized by an applied magnetic field, interact with conduction electrons. The localized magnetic moments polarize the angular orbital momentum of the conduction electrons, leading to a nonzero AHE[40]. The AHE vanishes in the absence of an applied magnetic field. The magnetic moments are typically point defects with localized unpaired electrons. Here, oxygen vacancies could play the role of paramagnetic centers.

Density Functional Theory (DFT) calculations with hybrid exchange-correlation functionals were performed to determine the magnetic property of the oxygen vacancies. Details of the DFT calculation method are described in the supplementary materials and were used previously to illustrate the point defects in other perovskite oxide systems[41,42]. These calculations identified that the $V_O^{1+}$ defect is stable when the Fermi level is near or above the conduction band minimum, as is the case here, and in this charge state the defect has a magnetic moment of ~1 $\mu_B$ (Supplementary materials, S4).

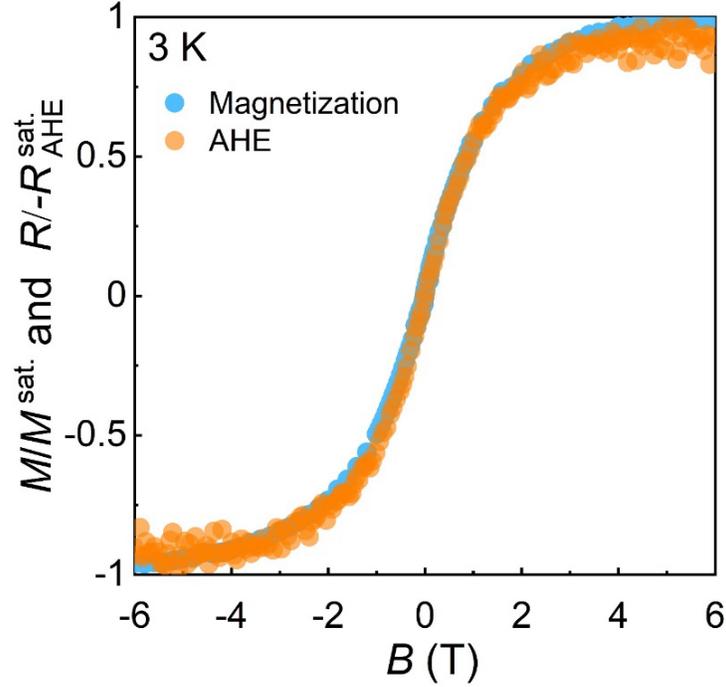

**Figure 3. Comparison of the AHE and magnetization in reduced KTaO$_3$.** The normalized AHE and magnetization responses in reduced KTaO$_3$ at 3K.

The AHE and magnetization should have similar trends with the magnetic field, $R_{xy}^{AHE}(B) = \lambda M_z(B)$ as shown in Fig. 3. The magnetization experiments at different temperatures show a paramagnetic signal in addition to the diamagnetic response in the reduced KTaO$_3$. The diamagnetic response was isolated from the high field results (5 T<$B$<7 T) and subtracted from the raw magnetization data (Supplementary materials S5). The Arrott plots confirm a paramagnetic response at 3 K (Supplementary materials, S6). The pristine KTaO$_3$ single crystals show a significantly weaker magnetic response (Supplementary materials, S7). The oxygen vacancy density is estimated from the Hall carrier density, and each oxygen vacancy shows ~1 $\mu_B$ using SQUID magnetometry, which is consistent with DFT calculations. The Hall carrier density, however, does not account for the trapped carriers. For example, it was recently demonstrated that oxygen vacancies form clusters which give rise to localized mid-gap states in KTaO$_3$[30]. Here, we speculate that oxygen vacancies act as paramagnetic centers with unpaired spins. We normalized the magnetization and AHE to their saturation value at high field to compare their trend. Figure 3 shows that the normalized magnetization and AHE are in close agreement, confirming that the magnetic nature of oxygen vacancies gives rise to AHE.

Figure 4(a) and (b) show temperature dependence of anomalous Hall saturation resistance and conductivity, $\sigma_{AHE} = R_{AHE}^{Sat.}/[R_{AHE}^{Sat.\,2} + R_{xx}^2(B=0)]$. The AHE saturation resistance shows an abrupt increase below 5 K. The anomalous Hall conductivity($\sigma_{AHE}$) shows an enhancement upon cooling and plateaus below 5 K. The insensitivity of Hall conductivity to longitudinal conductivity, i.e. scattering rate, below 5 K (Fig. 4(b) inset) is the signature of an intrinsic anomalous Hall effect in which Berry curvature of the conduction electrons on the Fermi surface gives rise to AHE. Figure 4(c) shows the dependence of the anomalous Hall conductivity to longitudinal conductivity above 5 K. Here, the results show a power law scaling ($\sigma_{xy}^{AHE} \propto \sigma_{xx}^{\alpha}$) with $\alpha \approx 1.6$, suggesting a suppression of intrinsic AHE by strong scattering of charge carriers. Screening of the scattering events in the low temperature limit triggers an extrinsic to intrinsic anomalous Hall *crossover*.

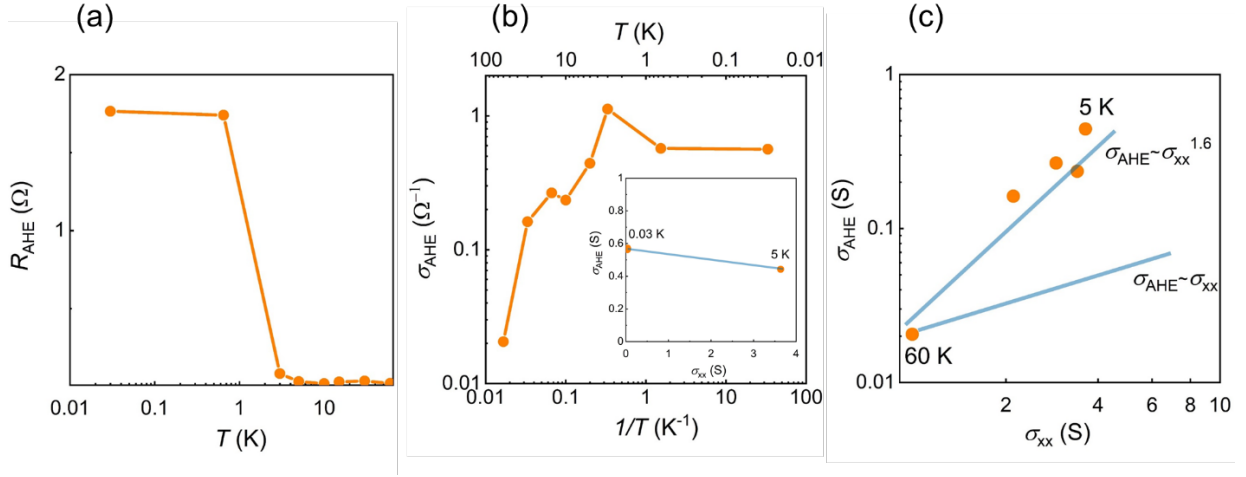

**Figure 4. The extrinsic to intrinsic anomalous Hall crossover**. (a) Anomalous Hall saturation resistance with temperature. (b) Anomalous Hall conductivity with temperature and longitudinal conductivity (inset). (c) Scaling relations between anomalous Hall conductivity and longitudinal conductivity shows a power law behavior ($\sigma_{xy}^{AHE} \propto \sigma_{xx}^{1.6}$) at higher temperatures (5-60 K).

**Discussion**

To briefly summarize the results, our main findings are as follows: (i) Reduced KTaO$_3$ shows a robust anomalous Hall response; (ii) AHE follows the oxygen vacancy-induced magnetization;

and (iii) The anomalous Hall response becomes non-dissipative below 5 K, suggesting an extrinsic to intrinsic crossover.

The first important conclusion from these results is that $KTaO_{3-\delta}$, a nominally non-magnetic system, shows a non-vanishing AHE down to 30 mK. The pristine crystals show significantly weaker magnetic response (Supplementary materials, S7), confirming that magnetic impurities do not control the magnetic response. Furthermore, the AHE could not be explained by a two-band fitting of the Hall results. Here, oxygen vacancies act as localized paramagnetic centers and give rise to AHE. DFT calculations and SQUID magnetometry resolve similar magnetic polarization values for oxygen vacancies. Surprisingly, the weak anti-localization correction to the longitudinal conductivity does not affect the scaling power factor. This could be attributed to the fact that only a fraction of charge carriers at the interface show weak anti-localization.

We observe a positive magnetoresistance, which is not expected in conventional AHE systems, i.e., itinerant magnets[31,43]. The Giovannini-Kondo (GK) model describes an extrinsic anomalous Hall effect based on coupling between orbital motion of the conduction electrons and localized magnetic moments[32,40]. This model, interestingly, does not require the conduction electrons to be magnetically polarized and is consistent with observed positive magnetoresistance. The GK picture, however, predicts a linear scaling ($\alpha = 1$) between longitudinal conductivity and AHE conductivity. As a result, the model describes the AHE in the clean limit due to skew scattering of charge carrier[44]. Here, the scaling power factor is ~1.6 at higher temperature regime (60-5 K). Decreasing the temperature suppresses the scattering events and the extrinsic to intrinsic crossover occurs when the relaxation rate becomes comparable to the Fermi energy, $h/\tau \propto E_F$, where $h$, $\tau$, and $E_F$ are Planck's constant, relaxation rate, and Fermi energy, respectively[22].

In summary, our results, especially the temperature dependence of the anomalous Hall effect, should be of interest for testing different theoretical models that have been proposed in the literature that relate scattering and AHE[4,5,7,21,45]. It would also be interesting to explore whether similar behavior will be observed in $KTaO_3$ 2D electron systems, especially at superconducting interfaces.

**Acknowledgements**
NCSU team was supported by the U.S. National Science Foundation under Grant No. NSF DMR-1751455. The authors acknowledge use of the SQUID and PPMS facility in the Department of Materials Science and Engineering at North Carolina State University.

**Data availability statement**

The data that support the findings of this study are available within the article.

**Conflict of Interests**

The authors declare that they have no conflict of interest.

# Supplementary information

## Oxygen vacancy-induced anomalous Hall effect in a non-magnetic oxide

Athby H. Al-Tawhid[1], Jesse Kanter[2], Mehdi Hatefipour[2], Douglas L. Irving[1,3], Divine P. Kumah[3], Javad Shabani[2], and Kaveh Ahadi[1,3,*]

[1]Department of Materials Science and Engineering, North Carolina State University, Raleigh, NC 27695, USA

[2]Center for Quantum Phenomena, Department of Physics, New York University, New York 10003, USA

[3]Department of Physics, North Carolina State University, Raleigh, North Carolina 27695, USA

**Magneto-transport results**

Transverse magnetoresistance was obtained at different temperatures using a Van der Pauw geometry. The temperature dependent magneto-transport measurements were performed using a Quantum Design physical property measurement system (PPMS) with a lock-in amplifier (SR830, Stanford Research Systems) in AC mode with an excitation current of 10 µA and a frequency of 13.33 Hz. The magneto-transport measurements below 1 K were carried out using a Tritondilution refrigerator (Oxford Instruments Group). The raw hall data was anti-symmetrized, $R_{xy} = [R_{xy}^{raw}(+B) - R_{xy}^{raw}(-B)]/2$ to eliminate the longitudinal magnetoresistance contribution. The linear component of the hall data was then extracted from the high field component where the AHE saturates, $R_H = dR_{xy}/dB$. The longitudinal magneto-transport was obtained by symmetrizing the raw magnetoresistance data, $R_{xy} = [R_{xy}^{raw}(+B) + R_{xy}^{raw}(-B)]/2$. The magneto-conduction was calculated by inverting the resistance tensor, $\sigma_{xy} = R_{xy}/[R_{xy}^2 + R_{xx}^2]$. The longitudinal magneto-conductivity (Fig. S2) shows a sharp peak at zero magnetic field, suggesting a weak anti-localization correction to conductivity. The coherence length of charge carriers must remain larger than conduction channel thickness, implying that only the electrons near the KTaO$_3$ surface show weak anti-localization.

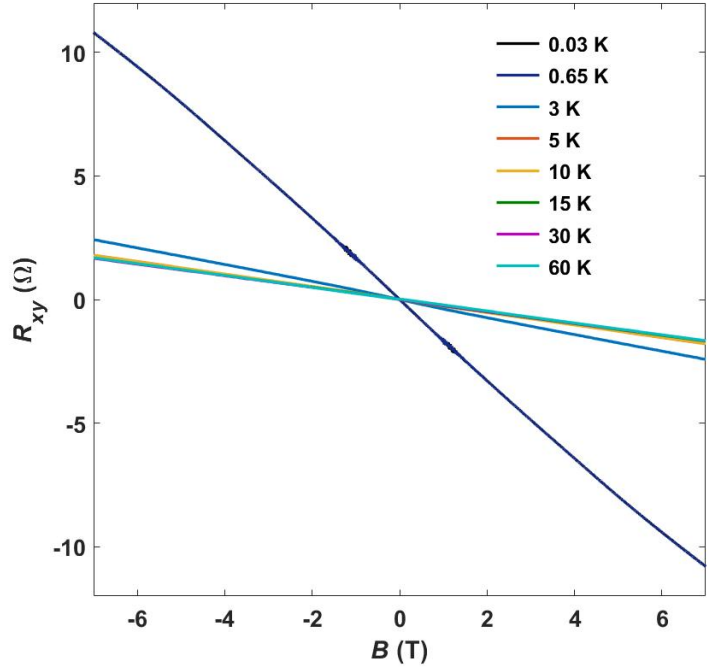

**Figure S1. Anti-symmetrized hall results at different temperatures.** The Hall results were anti-symmetrized to eliminate the longitudinal magneto-resistance contributions.

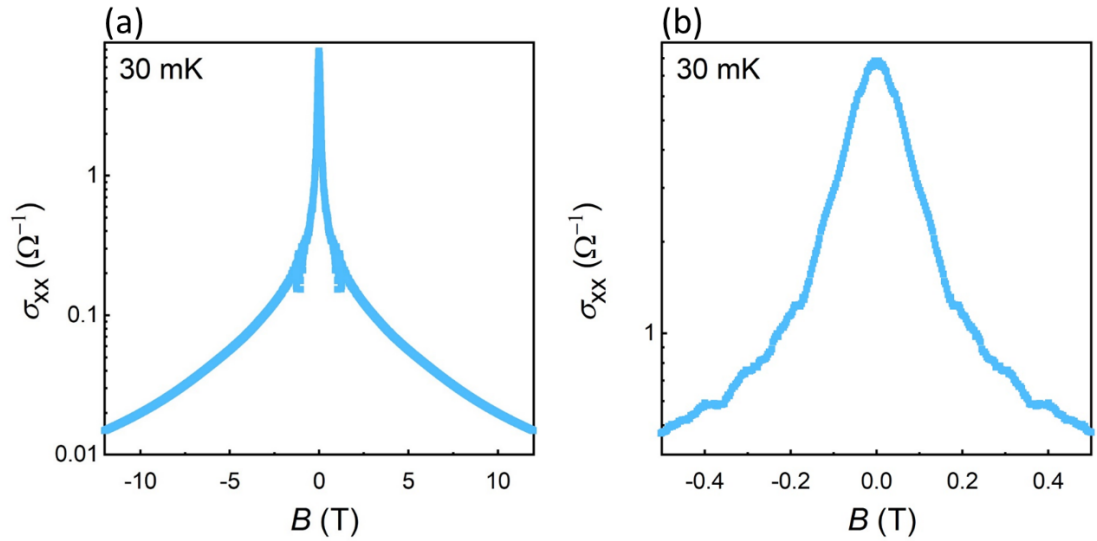

**Figure S2. Weak anti-localization in reduced $KTaO_3$.** (a) Magneto-conductance measurement of reduced $KTaO_3$ at 30 mK. (b) Zoom in magneto-conductance, showing a sharp peak at zero applied magnetic field, suggesting a weak anti-localization correction to conductivity.

**Two-band fit to Hall results**

The anti-symmetrized Hall results were modeled using a two-band model (Eq. S1).

$$R_{xy} = \frac{\mu_1^2 n_1 + \mu_2^2 n_2 + N \times (\mu_1\mu_2 B)^2}{e(\mu_1|n_1| + \mu_2|n_2|)^2 + N^2 \times (\mu_1\mu_2 B)^2} B \qquad (S1)$$

where μ is the mobility and n is the sheet carrier concentration of each band, $N$ is the total sheet carrier concentration, e is the elementary charge, and $B$ is the magnetic field. The fit was constrained by

$$N = n_1 + n_2 \qquad (S2) \qquad \text{and} \qquad \frac{1}{R_s} = e(\mu_1 n_1 + \mu_2 n_2) \qquad (S3)$$

The $n_1$ and $\mu_1$ are the only fitting parameters. The best fit, shown in Figure S3, does not describe the anti-symmetrized hall data.

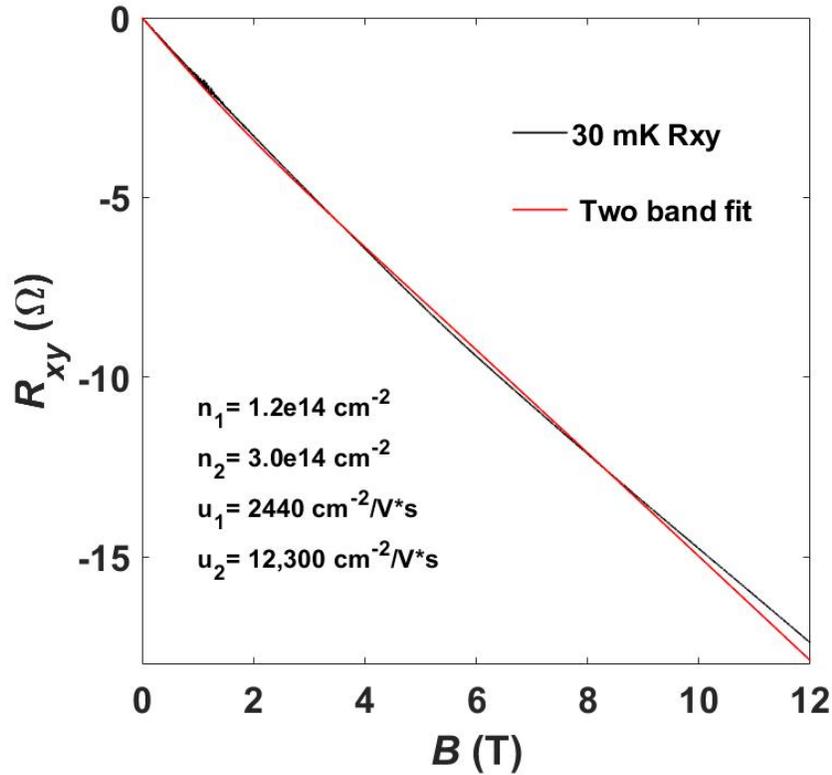

**Figure S3. Two-Band fit of the Hall results at 30 mK.** The fit of anti-symmetrized hall data at 30 mK using a two-band model.

**Calculation setup**

First principles density functional theory (DFT) calculations were performed using the Vienna Ab Initio Simulation Package (VASP) 5.[1–4] The screened hybrid exchange-correlation functional of Heyd, Scuseria, and Ernzerhof (HSE06)[5–7] with an exact exchange amount of 0.271 was used to reproduce the indirect (R-Γ) bandgap of 3.64 eV for KTaO$_3$. This fraction also improves the prediction of the optical gap (Γ-Γ), which is 4.29 eV with this approach and compares favorably to the experimental measurement of 4.35 eV.[8] Collinear spin polarization was accounted for in all calculations and a cut-off energy of 520 eV was used. Inclusion of spin-orbit coupling dramatically increases the expense of the calculations. Previous tests of inclusion led to only small shifts in prediction of thermodynamic transition levels (TTLs) for native defects and transition metal impurities. For this reason, they are not expected to change the qualitative trends predicted here but small changes to the TTL could be expected. Oxygen, potassium, and tantalum were simulated with 6, 9, and 11 valence electrons respectively. The 0 K lattice parameter for KTaO$_3$ using these parameters was 0.3987 nm. The oxygen vacancy was simulated in a 135 atom supercell and a 2×2×2 k-point mesh was used for the defect calculations. Atoms within a sphere of 5 Angstroms were free to relax with atoms outside this sphere held fixed in their ideal bulk positions. Finite sized corrections of the defects were performed using a modified approach of Kumagai and Oba.[9] Point defect information was managed with a point defects informatics framework.[10] These parameters, and their convergences, have been widely tested in complementary structures of SrTiO$_3$ and BaTiO$_3$.[11–16]

The formation energy of the oxygen vacancy was calculated within the established grand canonical formalsm.[17,18] The formation energy of defect D in charge state q is given in Eq. S4.

$$E_{D^q}^f = E_{D^q}^{tot} - E_{bulk}^{tot} - \sum_i n_i \mu_i + q\mu_e + E_{D^q}^{corr} \tag{S4}$$

In this expression, $E_{D^q}^f$ is the formation energy of defect $D$ in charge state $q$, $E_{D^q}^{tot}$ is the total energy of a supercell containing this defect in its respective charge state, $E_{bulk}^{tot}$ is the total energy of the ideal KTaO$_3$ bulk cell, $n_i$ atoms are exchanged with the chemical reservoir with chemical potential $\mu_i$ (here this is the chemical potential of oxygen), electronic charges are exchanged with the Fermi level $\mu_e$ that is used as a free parameter to plot the formation energies at 0 K and is taken relative to the valence band maximum (VBM), and $E_{D^q}^{corr}$ is the finite size correction discussed previously.

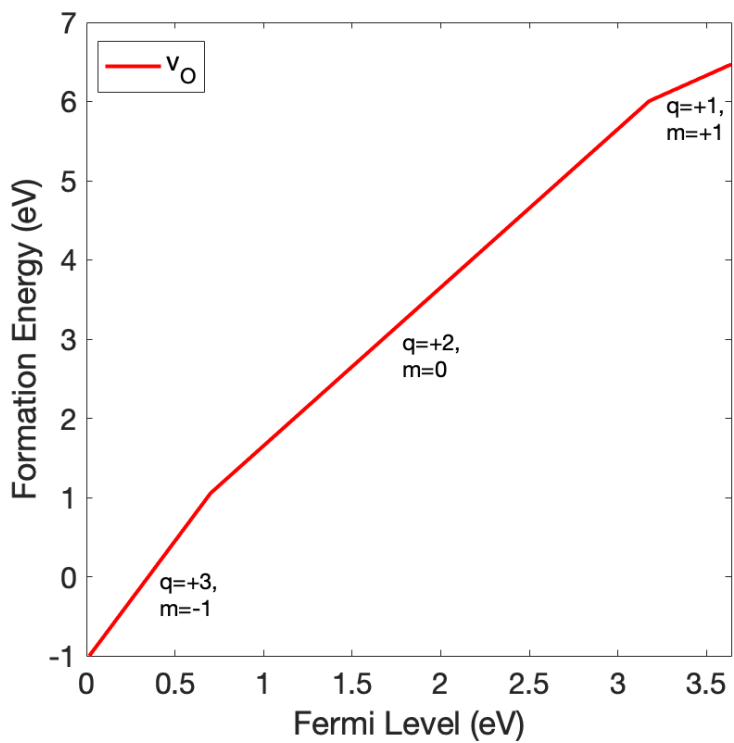

**Figure S4**: The formation energy of the oxygen vacancy ($v_O$) as a function of Fermi level. The stable charge states are labeled with changes in slopes indicative of the position of the TTLs. The magnetic moment associated with each charge state is also listed, which is obtained through sums of the occupied spin channels for the supercell containing the defect. The bulk itself has no associated moment.

The formation energy of the oxygen vacancy in oxygen rich conditions is plotted in Figure S4. The Fermi level in this plot is 0 at the VBM. The slope is indicative of the stable defect charge state for that particular Fermi level. The charge state is also labeled for convenience. Changes in slope are indicative of TTLs of this defect, or the ionization states of the oxygen vacancy. The ideal KTO in these calculations is non-magnetic. Also listed in the figures are the associated moments of the defect cells containing the oxygen vacancy. The $v_O^{2+}$ defect has no predicted magnetic moment and is stable in the central portion of the bandgap. For conductive samples studied here, the Fermi level would be near or above the conduction band minimum. In this portion of the bandgap, the stable oxygen vacancy is the $v_O^{1+}$, which has an associated magnetic moment of ~1 $\mu_B$, which is consistent to the presence of moments for the conductivity conditions measured experimentally. The formation energies are plotted at 0 K and in O-rich conditions and thus a maximum. Moving to reducing conditions would lower this curve but this would not change the TTLs. While this reduces the magnitude of the formation energy, it is still high as compared to that in SrTiO$_3$ and BaTiO$_3$, indicating that it is more difficult to form oxygen vacancies in KTaO$_3$ as compared to other perovskites.[13] This is consistent with the strength of the Ta-O bonds in KTaO$_3$.

**Magnetization response in reduced KTaO$_3$**

The magnetic properties of the reduced KTaO$_3$ were measured at different temperatures using a Quantum Design superconducting quantum interference device (SQUID) with a quartz holder. A diamagnetic behavior dominates the magnetic response of the sample, however, a small paramagnetic signal was found with subtracting the diamagnetic response. The paramagnetic response was obtained by removing the high field contribution where the paramagnetic response saturates. Figure S6 shows the Arrott plot at 3 K, confirming the paramagnetic nature of magnetic response at helium liquid temperature using a simple mean field theory. The pristine samples show a significantly smaller magnetization response (Figure S7). We conclude that oxygen vacancies are the main source of paramagnetic response.

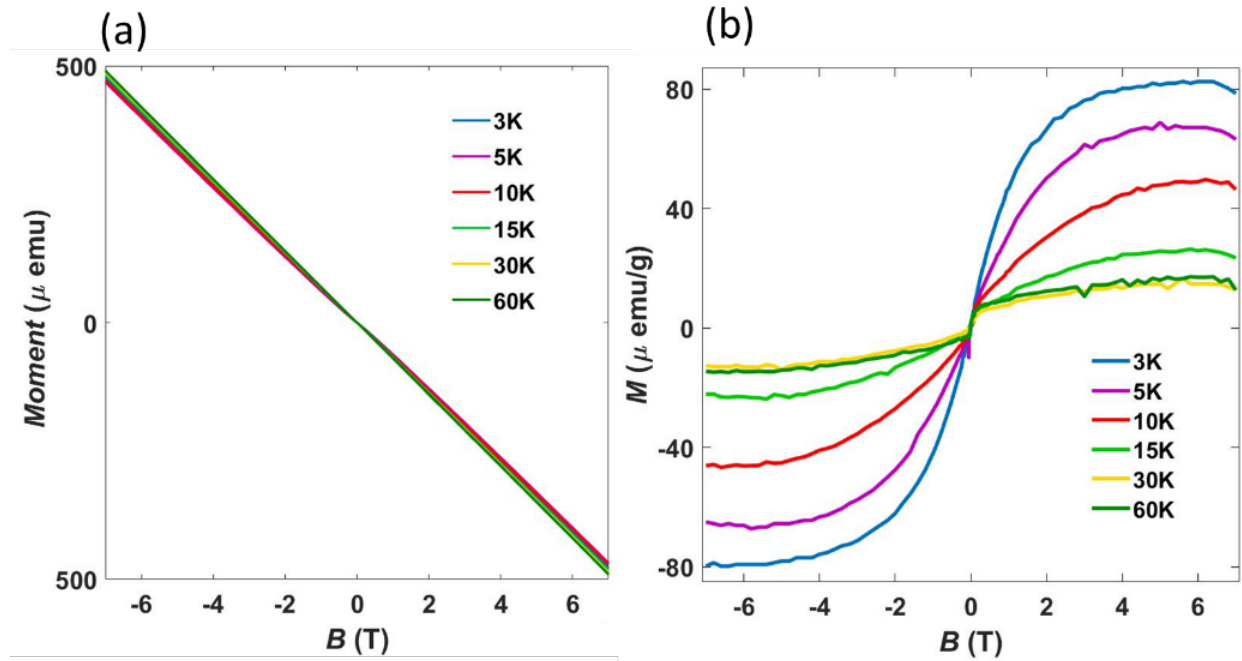

**Figure S5. Paramagnetic response of reduced KTaO$_3$.** (a) The raw magnetic moment with applied magnetic field measured at different temperatures. (b) Oxygen vacancy-induced paramagnetic response in reduced KTaO$_3$.

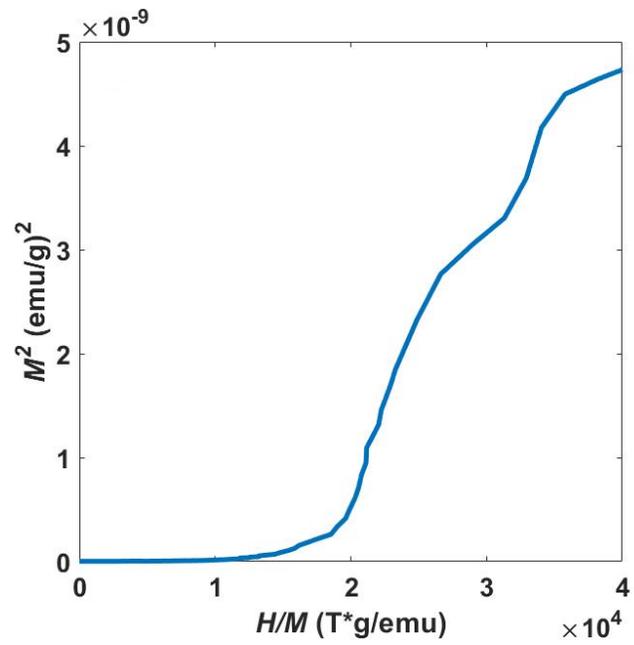

**Figure S6. Paramagnetic response of reduced KTaO$_3$.** Arrott plot using the magnetization curve of reduced KTaO$_3$ at 3 K, confirming a paramagnetic behavior.

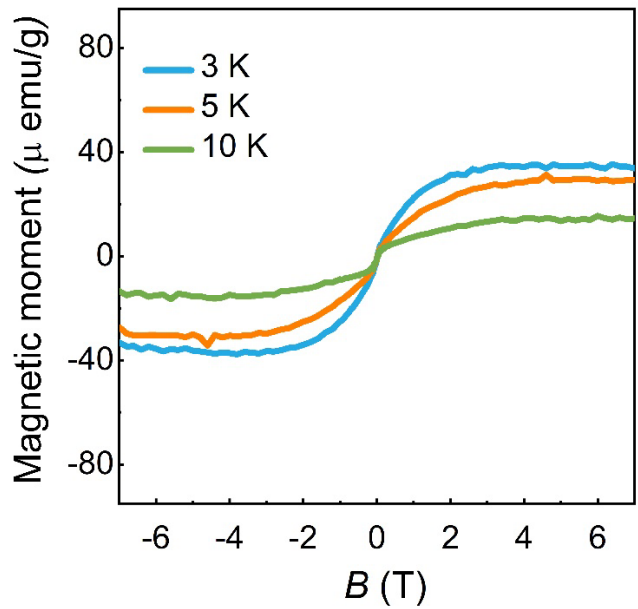

**Figure S7. Paramagnetic response of pristine KTaO$_3$ substrate.** The pristine substrates show a significantly smaller magnetic moment compared to reduced substrates.